\documentclass[12pt]{iopart}
\usepackage{iopams}
\usepackage{setstack}
\usepackage{hyperref} 
\usepackage{bm}
\hypersetup{colorlinks=true, linktoc=all, linkcolor=blue}
\newcommand{\avg}[1]{\langle #1\rangle}

\begin{document}

\title[Parameterized Post-Newtonian Cosmology]{Parameterized Post-Newtonian Cosmology}
\author{Viraj A A Sanghai$^1$ and Timothy Clifton$^2$}
\address{School of Physics \& Astronomy,
Queen Mary University of London, London, UK}
\ead{$^1$v.a.a.sanghai@qmul.ac.uk, $^2$t.clifton@qmul.ac.uk}

\begin{abstract}
Einstein's theory of gravity has been extensively tested on solar system scales, and for isolated astrophysical systems, using the perturbative framework known as the parameterized post-Newtonian (PPN) formalism. This framework is designed for use in the weak-field and slow-motion limit of gravity, and can be used to constrain a large class of metric theories of gravity with data collected from the aforementioned systems. Given the potential of future surveys to probe cosmological scales to high precision, it is a topic of much contemporary interest to construct a similar framework to link Einstein's theory of gravity and its alternatives to observations on cosmological scales. Our approach to this problem is to adapt and extend the existing PPN formalism for use in cosmology. We derive a set of equations that use the same parameters to consistently model both weak fields and cosmology. This allows us to parameterize a large class of modified theories of gravity and dark energy models on cosmological scales, using just four functions of time. These four functions can be directly linked to the background expansion of the universe, first-order cosmological perturbations, and the weak-field limit of the theory. They also reduce to the standard PPN parameters on solar system scales. We illustrate how dark energy models and scalar-tensor and vector-tensor theories of gravity fit into this framework, which we refer to as ``parameterized post-Newtonian cosmology'' (PPNC). 

\end{abstract}


\maketitle

\section{Introduction}

Einstein's theory of gravity has now been tested extensively in the solar system and in binary pulsar systems, using a wide array of relativistic gravitational phenomena \cite{lr}. These range from the deflection of light \cite{cassini}, to perihelion precession \cite{perihelion}, geodetic precession \cite{gravityprobe}, and frame dragging \cite{gravityprobe}. In all cases, the standard framework that is used to interpret observations of these effects is the parameterized post-Newtonian (PPN) formalism \cite{Will}. This formalism is constructed so that it encompasses the possible consequences of a wide variety of metric theories of gravity, and so that it can act as a half-way house between the worlds of experimental and theoretical gravitational physics. The PPN formalism has been tremendously successful not just in constraining particular modified theories of gravity, but also in providing a common language that can be used to isolate and discuss the different physical degrees of freedom in the gravitational field. Crucially, the form of the PPN metric is independent of the field equations of the underlying theory of gravity, and is simple enough to be effectively constrained with imperfect, real-world observations. It is also applicable in the regime of non-linear density contrasts. These are all highly desirable properties.

With the advent of a new generation of cosmological surveys \cite{euclid, ska, lsst}, it becomes pertinent to consider whether we can perform precision tests of Einstein's theory on cosmological scales. Of course, the standard PPN formalism itself cannot be used directly for this purpose, as it is valid only for isolated astrophysical systems. More specifically, it relies on (i) asymptotical flatness and (ii) the  slow variation of all quantities that might be linked to cosmic evolution. Neither of these conditions should be expected to be valid when considering gravitational fields on large scales: there are no asymptotically flat regions in cosmology, and the time-scale of cosmic evolution is no longer necessarily entirely negligible. We must therefore adapt and extend the PPN approach, if it is to be used in cosmology. Some of this work has already been performed within the context of Einstein's theory \cite{Tim1,vaas1, vaas2}, but more is required if we are going to attempt to port the entire formalism. This is what we intend to make a step towards in the present paper, in a formalism that we will call parameterized post-Newtonian cosmology (PPNC). 

Of course, we wish to retain as many of the beneficial properties of the PPN formalism as possible. In particular, we want to ensure that the formalism is still valid in the presence of non-linear structure after it has been transferred into cosmology. We also want to ensure that it can encompass as large a class of theories of gravity and dark energy models as possible, while remaining simple enough to be constrained by real observations. These requirements are important as many cosmological processes take place in the presence of non-linear structures, and because we want to be able to represent as many theories as possible. The parameterization that we end up with contains four functions of time that we expect to be able to link to the large-scale expansion, the growth of structure, and the lensing of light in a reasonably straightforward way. We do not assume any knowledge of the specific underlying theory of gravity in order to end up with this result, other than insisting that it fits into the class of conservative theories that can be described using the PPN formalism. Our approach is built using a weak-field and slow-motion post-Newtonian expansion, and so is naturally valid in the presence of non-linear structures (up to neutron star densities).

Our work builds on a series of bottom-up approaches to cosmology, that has so far primarily been usually used to study the effect of small-scale inhomogeneities on the large-scale expansion within the context of general relativity \cite{Tim1}-\cite{Tim5} (exceptions to this are applications to $f(R)$ gravity \cite{Tim2b,Tim2} and Yukawa gravity \cite{pierre1}). We also expect our study to complement the existing literature on parameterized frameworks for testing gravity in cosmology, which come under the umbrella terms of ``parameterized post-Friedmannian'' approaches \cite{Hu1}-\cite{PPF3} and ``effective field theory'' approaches and their variants \cite{eff1}-\cite{eff8}. Our approach differs from most of this existing literature in the fact that we emphasize the links between weak gravitational fields and cosmology, and use this to constrain the possibilities for the large-scale properties of cosmology. This means that we end up with a framework that is automatically consistent with the PPN formalism on small scales, and that is constrained by this consistency in the form that it can take on large scales. For reviews on modified theories of gravity and parameterized frameworks in cosmology, the reader is referred to Refs. \cite{Tim3} and \cite{Tim3b}.

The plan for the rest of this paper is as follows. In Section \ref{sec2} we introduce the bottom-up constructions we will use to link weak-field gravity and cosmology \cite{Tim1,vaas1, vaas2}. Section \ref{sec3} contains a review of the standard parameterized post-Newtonian approach, which we then modify for application to cosmology. In Section \ref{cosmo}, we build a cosmology from the weak field metric without assuming any field equations. This results in a geometry with four unknown functions of time. Finally, in Section \ref{examples}, we work through four explicit example theories, to show how we expect our formalism to function. Our examples include dark energy models, and scalar-tensor and vector-tensor theories of gravity. We use lower-case Latin letters ($a$, $b$, $c$, ...) to denote space-time indices, and Greek letters ($\mu$, $\nu$, $\rho$, ...) to denote spatial indices. Capitals from the first half of the Latin alphabet ($A$, $B$, $C$, ...) are used to denote the spatial components of tensors in $1+2$-dimensional subspaces, while those from the latter half ($I$, $J$, $K$, ...) will be used to label quantities associated with various different matter fields.

\section{From weak fields to cosmology} \label{sec2}

In this section we wish to explore the relationship between weak-field gravity and cosmology, without assuming anything about the field equations that govern the gravitational interaction (i.e. without assuming a specific theory of gravity). These two sectors are usually treated entirely separately in the standard approach to cosmology, as they appear at different orders in cosmological perturbation theory. They are, however, intimately linked, and given some knowledge about the weak-field limit of gravity one can construct cosmological evolutions that are consistent with that limit. We do not require a set of field equations in order to do this, as long as we are considering metric theories of gravity. The end result is then a set of effective Friedmann equations in which the large-scale expansion is driven by sources that can be expressed in terms of weak-field potentials. The link between these potentials and the energy-momentum content of the universe can subsequently be determined by the particular field equations of the theory that one wishes to consider. The great benefit of writing the Friedmann equations in this way is that they can be directly expressed in terms of (an extended version of) the PPN parameters. This facilitates both a direct comparison of cosmological and weak-field tests of gravity, as well as constraining the otherwise near limitless freedoms that can exist when parameterizing gravitational interactions in cosmology.

\subsection{Post-Newtonian expansions}
\label{pn}

The perturbative approach we intend to use is the method of post-Newtonian expansions. This approach is designed to be applied to the weak-field and slow-motion limit of gravitational interactions, and is formally an expansion around Minkowski space in the parameter
\begin{equation}  \label{2}
\epsilon \equiv \frac{|\bm{v}|}{c} \ll 1 \, , 
\end{equation} 
where $c$ is the speed of light, and $\bm{v}$ is the three-velocity associated with matter fields. We can then use $\epsilon$ to assign orders of magnitude to the matter content and the metric perturbations, such that
\begin{equation}
\rho \sim \varphi \sim v^2 \sim \epsilon^2 \, ,
\end{equation}
where $\rho$ is the mass density, and $\varphi$ represents a generic gravitational potential. The post-Newtonian expansion is valid in the quasi-static regime, where time derivatives are small compared to space derivatives, such that
\begin{equation}
\frac{{\partial}/{\partial t}}{{\partial}/{\partial x}} \sim \epsilon \, .
\end{equation}
This means that the length scales associated with these gravitational fields must be small compared to the horizon size. We will therefore use the post-Newtonian expansion to describe small regions of space, and patch these regions together to determine the emergent large-scale cosmological expansion. For further details of post-Newtonian perturbative expansions the reader is referred to Ref. \cite{Will}.

\subsection{Expanding and non-expanding coordinate systems}

The PPN formalism, and post-Newtonian expansion generally, are formulated as an expansion about Minkowski space, such that the geometry can be described to lowest non-trivial order by
\begin{eqnarray}
ds^2 = -(1-2\Phi) dt^2 + (1+ 2\Psi)  (dx^2+dy^2+dz^2)  \, , \label{weakfield}
\end{eqnarray}
where the gravitational potentials are of order $\Phi \sim \Psi \sim \epsilon^2$. In the present context, it is useful to transform this line-element so that it can be written as a perturbed Friedmann geometry. The coordinate transformations required for this are \cite{Tim1, vaas1, vaas2}
\begin{eqnarray}
t &= \hat{t} + \frac{\dot{a} a}{2} (\hat{x}^2 + \hat{y}^2 + \hat{z}^2) + O(\epsilon^3) \ , \label{timetrans} \\[5pt]
x &= a \hat{x} \bigg[1 + \frac{\dot{a}^2}{4} (\hat{x}^2 + \hat{y}^2 + \hat{z}^2)\bigg] +O(\epsilon^4) \ , \label{xtrans}  \\[5pt]
y &= a \hat{y} \bigg[1 + \frac{\dot{a}^2}{4} (\hat{x}^2 + \hat{y}^2 + \hat{z}^2)\bigg] +O(\epsilon^4) \ , \\[5pt]
z &= a \hat{z} \bigg[1 + \frac{\dot{a}^2}{4} (\hat{x}^2 + \hat{y}^2 + \hat{z}^2)\bigg] +O(\epsilon^4) \label{ztrans} \, ,   
\end{eqnarray}
where {$a=a(\hat{t}) \sim O(1)$ and $\dot{a}=da(\hat{t})/d\hat{t} \sim O(\epsilon)$, because time derivatives add an order of smallness}. Applying these coordinate transformations to the perturbed Minkowski space in Eq. (\ref{weakfield}) gives, to lowest non-trivial order,
\begin{equation}
ds^{2} = -(1 - 2\hat{\Phi})d\hat{t}^2 +  a(\hat{t})^2 (1+2\hat{\Psi}) \frac{ \left( d \hat{x}^2 + d \hat{y}^2 + d \hat{z}^2 \right)}{[1+\frac{k}{4} (\hat{x}^2 + \hat{y}^2 + \hat{z}^2)]^2}  \, ,  \label{FLRW}
\end{equation}
where $\hat{\Phi}$ and $\hat{\Psi}$ are defined, up to terms of $O(\epsilon^4)$, by
\begin{eqnarray}
\Phi &=& \hat{\Phi} + \frac{\ddot{a} a}{2} (\hat{x}^2 + \hat{y}^2 +\hat{z}^2) \ , \label{phitran}\\[5pt]
\Psi &=& \hat{\Psi} - \bigg(\frac{\dot{a}^2 + k}{4}\bigg) (\hat{x}^2 + \hat{y}^2 +\hat{z}^2) \label{psitran}\, .   
\end{eqnarray}
The quantity $k \sim \epsilon^2$, that appears in (\ref{psitran}), is the Gaussian curvature of the conformal 3-space. The geometry and coordinate system used  (\ref{FLRW}) look identical to those of a global FLRW model with linear scalar perturbations. This is, however, only a coordinate transformation of the perturbed Minkowski space from equation (\ref{weakfield}). It is therefore only valid within the same region of space that the perturbed Minkowski description was valid (i.e. a space much smaller than the size of the horizon). The scale factor, $a(\hat{t})$, is not yet the solution to any set of Friedmann equations, and does not yet correspond to the scale factor of any global Friedmann space. It is simply an arbitrary function of time, introduced by the coordinate transformations in equations (\ref{timetrans}) - (\ref{ztrans}). In order to associate it with a global scale factor, and determine the relevant Friedmann equations, we must patch together many such regions of space, using appropriate junction conditions.

\subsection{Junction conditions}

The conditions required at the junction between neighbouring regions of space, in order for their union to be considered a solution of the field equations, will now be determined. Let us first choose to consider junctions that are $(2+1)$-dimensional time-like submanifolds of the global space-time. In this case, the space-like unit vector normal to the junction is given as the solution to
\begin{equation}
n_{a}\frac{ \partial x^{a}}{\partial \xi^{i}} = 0 \qquad {\rm and} \qquad n_a n^a =1 \, ,
\end{equation}
where $\xi^{i}$ are the coordinates on the boundary. The first and second fundamental forms on the boundary are then given by
\begin{equation}
\gamma_{ij}  = \frac{ \partial x^{a}}{\partial \xi^{i}} \frac{ \partial x^{b}}{\partial \xi^{j}} \gamma_{ab} \qquad {\rm and} \qquad K_{ij}\equiv \frac{1}{2} \frac{\partial x^{a}}{\partial \xi^{i}} \frac{\partial x^{b}}{\partial \xi^{j}} \mathcal{L}_{n} \gamma_{ab} \, ,
\end{equation}
where $\gamma_{ab} = g_{ab} - n_a n_b $ is the projection tensor onto the boundary. For a metric theory of gravity, we will expect to be able to impose certain conditions on the values of $\gamma_{ij}$ and $K_{ij}$, on either side of the junction.

Strictly speaking, the junction conditions on the geometry will depend on the specific field equations that apply to the theory of gravity that is being considered. However, it is reasonable to expect that certain junction conditions should result generically from conservatively constructed metric theories. In particular, we expect that the Israel junction conditions in the absence of surface layers should be obeyed. These conditions are given by \cite{Israel}
\begin{eqnarray}
\bigg[\gamma_{ij}\bigg]^{(+)}_{(-)} = 0 \ , \label{metjunc1} \\[5pt]
\bigg[K_{ij}\bigg]^{(+)}_{(-)} = 0 \, . \label{metjunc2}
\end{eqnarray}
where $[\varphi]^{(+)}_{(-)} = \varphi^{(+)} - \varphi^{(-)}$ for any object $\varphi$, and where superscripts $(+)$ and $(-)$ indicate that a quantity should be evaluated on either side of the boundary. {The first junction condition (\ref{metjunc1}) comes from the assumption of a continuous induced metric. This is both natural and required so that no Dirac delta functions arise while computing the affine connection. The second junction condition (\ref{metjunc2}) comes from the Ricci equation,}
\begin{eqnarray}
R_{ij} = R^{(3)}_{ij} + 2 K_{im}K^{m}_{\ j} - K_{ij}K^{m}_{\ m} - \mathcal{L}_{n} K_{ij} + \dot{n}_{(i;j)} \ ,
\end{eqnarray}
{where $R_{ij}$ is the Ricci curvature of space-time projected on the boundary, $R^{(3)}_{ij}$ is the Ricci curvature of the (2+1)-dimensional surface and $\dot{n}_{i} \equiv n_{i;b}n^{b}$. If $K_{ij}$ was discontinuous we would have a Dirac delta function in the $\mathcal{L}_{n} K_{ij}$ term, and hence also in the Ricci curvature. Generically, we expect the Ricci curvature to be related to the energy-momentum tensor, in any theory of gravity that contains second derivatives of the metric in the field equations. This means that if Eq. (\ref{metjunc2}) were not satisfied then we would generically expect to have a discontinuity in the energy-momentum tensor. However, as we are considering situations where there are no surface layers or branes on the boundary, this is not something that can be allowed. We therefore expect the junction conditions (\ref{metjunc1}) and  (\ref{metjunc2}) to apply to any covariant theory of gravity that contains second derivatives of the metric in its field equations, as they simply correspond to the metric being $C^1$ smooth at the boundary. This expectation has shown to hold true in scalar-tensor theories \cite{scalarjunc} and $f(R)$ theories of gravity \cite{sasaki}.} If they were found to be untrue, for any particular theory of gravity, then the theory in question would not fall into the domain of applicability of the framework we are constructing. Such anomalous theories would then have to be treated separately, as special cases.

The junction conditions (\ref{metjunc1}) and (\ref{metjunc2}) are sufficient to allow us to evaluate the motion of the boundaries of each of our small regions of space, and therefore also tell us the cosmological expansion we expect to obtain from regions described by the geometry in (\ref{weakfield}) and (\ref{FLRW}). This will be described in terms of the potentials $\Phi$ and $\Psi$ in Section \ref{emergent}, and in terms of (an extended set of) the PPN parameters in Section \ref{cosmo}. In Section \ref{examples} we will use these junction conditions, along with additional conditions where required, to relate the weak field geometry to the cosmological expansion in some specific example classes of modified theories that contain additional scalar and vector degrees of freedom. This will allow us to write the functions that appear in the Friedmann equation in terms of the parameters of these example theories.

\subsection{Emergent cosmological expansion}
\label{emergent}

The junction condition in (\ref{metjunc2}) is satisfied if $K_{ij}=0$, on the boundary of every small region of space. This condition means that the boundary is extrinsically flat in the $3+1$-dimensional space-time, and is probably the simplest way of satisfying the second junction condition. Examples of constructions with time-like boundaries of this type are the regular lattices of discrete masses studied in Refs. \cite{Tim1}-\cite{Tim5}, but it is also a perfectly good way to describe an FLRW space that has been divided into small sub-regions with comoving flat boundaries. If we choose to consider regions of space with extrinsically flat boundaries of this type,  then we find that this implies \cite{Tim1,vaas1,vaas2}
\begin{eqnarray} 
{{X}_{,tt}} &=& \mathbf{n}\cdot\nabla\Phi |_{\partial \Omega} + O(\epsilon^4) \, , \label{X1}  \\[5pt] 
X_{,AB} &=& \delta_{AB} \, \mathbf{n}\cdot\nabla\Psi|_{\partial \Omega}  + O(\epsilon^4) \, , \label{X2} \\[5pt]
X_{,tA} &=& 0 + O(\epsilon^3)  \ . \label{X3}
\end{eqnarray}
where we have rotated coordinates so the boundary is located at $x=X(t,y,z)$ (to first approximation). The $|_{\partial \Omega}$ symbol in this equation indicates that the preceeding quantity is being evaluated on the boundary of the region under consideration. These equations describe the motion of the boundary of our small region of space, as well as its shape. After transforming to expanding coordinates via equations (\ref{timetrans}) - (\ref{ztrans}), and choosing $a(t)$ such that each part of the boundary is comoving with the $(\hat{x},\hat{y},\hat{z})$ coordinates, we can use equation (\ref{X1}) to write one of the Friedmann equations for the global space. This will be explained further in Section \ref{cosmo}, after introducing the relevant formalism in Section \ref{sec3}.

The other Friedmann equation requires us to derive a Hamiltonian constraint equation. To do this we again assume that there exists a coordinate system where every part of the boundary is comoving with the $(\hat{x},\hat{y},\hat{z})$ coordinates, and consider a time-like 4-vector field that is both uniformly expanding and comoving with our boundaries:
\begin{equation} \label{ua}
u^a = \left(1 ; \frac{X_{,t}}{X} x^{\mu} \right) \, ,
\end{equation}
where we have kept only the leading-order term in each component, and where we have expressed the components in the $(t,x,y,z)$ coordinates. A spatial hyper-surface orthogonal to this field then gives, from a post-Newtonian expansion of Gauss' embedding equation, that
\begin{equation}
\left( \frac{X_{,t}}{X} \right)^2 = -\frac{2}{3} \nabla^2 \Psi - \frac{R^{(3)}}{6}  + O(\epsilon^4) \, , \label{con1}
\end{equation}
where $R^{(3)}$ is the Ricci curvature scalar of the space, which for the situation we are considering can be related to the spatial curvature, $k$. The functional form of equation (\ref{con1}) is strongly reminiscent of the Friedmann equations, and after transformation to the expanding coordinates can also be used to construct an effective Friedmann equation for the global space. Again, this will be explained further in Section \ref{cosmo}.

We emphasize that nowhere in this section have we assumed anything about any theory of gravity or a set of field equations, other than the junction conditions (\ref{metjunc1}) and (\ref{metjunc2}). Nevertheless, we have ended up with a set of equations that looks very similar to the Friedmann equations, with sources given by the derivatives of weak-field potentials. A concrete realisation of the types of structure being described here is a regular lattice, constructed from cells that are themselves regular convex polyhedra. Such structures were considered in the context of Einstein's theory in Refs. \cite{Tim1, vaas1,vaas2}, and will often be what we have in mind in what follows.

\section{An extended PPN formalism} \label{sec3}

Let us now consider how to extend the PPN framework, so that it can be used to model weak gravitational fields in an expanding universe. We will begin by briefly discussing the basics of the existing PPN formalism, as it is currently found in the literature \cite{Will}. We will then discuss how we can extend it to include other forms of matter that are relevant in cosmology, and to include the time dependence that is a crucial feature of an expanding universe. This will require not only allowing the parameters themselves to be dynamical, but also the boundary conditions that we use for solving the relevant hierarchy of Poisson equations.

\subsection{The standard PPN formalism}

The standard PPN formalism is built upon the post-Newtonian expansions outlined in Section \ref{pn}. It does not assume any particular form for the field equations, but does make an ansatz for the weak field metric (which is expected to be valid for any metric theory of gravity).  Up to $O(\epsilon^2)$, this PPN metric is given by equation (\ref{weakfield}), which has already been written in the standard post-Newtonian gauge, so that it is diagonal and isotropic at leading order in perturbations. As well as the metric, the energy-momentum tensor is also subject to a post-Newtonian expansion. To lowest non-trivial order, this gives
\begin{eqnarray}
T_{tt}=& \rho_{M}(t, x^{\mu}) + O(\epsilon^4) \label{emPPNtt} \ ,  \\[5pt]
T_{t\mu} =& - \rho_{M}(t, x^{\mu}) \, v_{M\mu}(t, x^{\mu}) + O(\epsilon^5) \label{emPPNtx} \ , \\[5pt]    
T_{\mu \nu} =& p_M(t, x^{\mu}) \delta_{\mu \nu} + O(\epsilon^6) \ , \label{emPPN}
\end{eqnarray}
where $\rho_{M}(t, x^{\mu})\sim \epsilon^2$ is the mass density of non-relativistic matter, $v_{M\mu}(t, x^{\mu})\sim \epsilon$ is the 3-velocity of this matter, and $p_M(t, x^{\mu})\sim \epsilon^4$ is the isotropic pressure. This energy-momentum tensor is assumed to be conserved, so that $T^{ab}_{\ \ ; a} =0$.

The relationship between gravitational potentials and energy-momentum content is, of course, specified by the gravitational field equations. If these equations are unknown, or we do not want to specify any particular theory of gravity, then the best we can do is simply assume that the Laplacian of the gravitational potentials can be expressed as a linear function of the energy-momentum content of the space-time. This is done in the PPN framework by writing\footnote{The usual definition of $\alpha$ and $\gamma$ actually involves the solution to this equation written in terms of the integrals of an asymptotically flat Green's function. We have presented it in this way so that it is more amenable for adaption to cosmology.} 
\begin{eqnarray}
\nabla^2 \Phi = -4\pi G \alpha \rho_{M}  \, , \label{alpha}\\[5pt] \label{gamma} 
\nabla^2 \Psi = -4\pi G \gamma \rho_{M}
\end{eqnarray}
where $G$ is Newton's gravitational constant, and where $\alpha$ and $\gamma$ are constants.  {Of course, this description only applies to theories of gravity where Yukawa potentials are either absent, neglected, or can be approximated by Coulomb-like potentials. It also relies on the absence, or neglect, of any non-perturbative physics.} Inclusion of these types of gravitational interactions would require extending both the PPN framework, and the PPNC that we construct here.

Now, the lowest-order equations of motion for time-like particles, from $T^{ab}_{\ \ ; a} =0$, tells us that $\Phi$ is the gravitational potential that causes acceleration due to the Newtonian part of the gravitational field. For agreement with local experiments (i.e. so that $G$ is the locally measured value of Newton's constant), we must therefore have $\alpha=1$ at the present time. The parameter $\gamma$ then parameterizes the relativistic deflection of light and Shapiro time delay, while further constants (not given explicitly here) parameterize the zoo of other relativistic effects that are observable in the solar system and elsewhere. The current best observational constraints on this parameter are $\gamma = 1 + (2.1 \pm 2.3) \times 10^{-5}$ \cite{cassini}, which is consistent with the value $\gamma=1$ that is expected from Einstein's theory.

The description given above is sufficient to calculate the leading-order gravitational effects on both null and time-like particles. However, if we want to calculate explicit expressions for $\alpha$ and $\gamma$, in terms of the parameters of a given theory of gravity, then we must also expand the additional degrees of freedom present in that theory. For an additional scalar field, $\phi$, this expansion is usually taken to be
\begin{equation}
\label{s1}
\phi = \bar{\phi} + \delta \phi(t, x^{\mu}) + O(\epsilon^4) \, ,
\end{equation}
where $\bar{\phi}  \sim \epsilon^0$ is the constant background value of the scalar field, and where $\delta \phi(t,x^{\mu}) \sim \epsilon^2$ is the leading-order perturbation. Similarly, for a theory with a time-like vector field $A_{a}$, one can expand its components as 
\begin{eqnarray}
A_{t} = \bar{A}_{t} + \delta A_{t}(t, x^{\mu}) +O(\epsilon^4)\, , \label{v1} \\[5pt] \label{v2}
A_{\mu} = \delta A_{\mu}(t, x^{\mu}) +O(\epsilon^5)\, ,
\end{eqnarray}
where $\bar{A}_{t} \sim \epsilon^0$ is the background value of the time-component,  and $\delta A_{t}(t, x^{\mu}) \sim O(\epsilon^2)$ and $ \delta A_{\mu}(t, x^{\mu}) \sim O(\epsilon^3)$ are the leading-order perturbations to the time and space components of the vector field, respectively.

Of course, other types of additional fields can be included, depending on the types of theory that one wishes to consider. For further details on this, and other aspects of the standard PPN formalism, the reader is referred to Ref. \cite{Will}. In the following sections we will extend the PPN formalism by adding additional matter content, additional gravitational potentials, and by allowing for additional time dependence in the parameters. These extensions are all required in order to adapt the PPN formalism for cosmology.

\subsection{Additional matter content}

The treatment above assumes $p \ll \rho$, which is fine for the contents of the solar system, but for cosmological studies would confine us to considering dust. We would like our formalism to also be able to incorporate generic dark energy fluids, radiation, scalar fields, and the variety of other types of matter that are often studied in cosmology. We therefore take the total energy-momentum tensor of all matter fields to be given by
\begin{eqnarray}
T^{ab} = T^{ab}_M + \sum_{I} T^{ab}_I\ ,
\end{eqnarray}
where subscript $M$ refers to quantities associated with non-relativistic pressureless matter fields (i.e. baryons and dark matter), and where subscript $I$ refers to quantities associated with all other barotropic fluids. The energy-momentum tensor of each of these fluids can then be written
\begin{equation}
T_J^{a b} = \rho_J u_J^a u_J^b + p_J (g^{ab} +u_J^a u_J^b) \, ,
\end{equation}
where we intend $J \in \{M,I\}$, and where the 4-velocity $u_J^a$ can be written
\begin{eqnarray} 
u_J^{a} =  \bigg(1 +\Phi + \frac{v_J^2}{2}\bigg)(1;v_J^{\mu}) +  O(\epsilon^4) \ , \label{4-vel1}
\end{eqnarray}
where $v_J^{\mu}$ is the 3-velocity of fluid $J$, and where $v_J^2= v_J^{\mu}v_{J \mu}$. The components of the total energy-momentum tensor are then given, to leading order, by
\begin{eqnarray}
T_{tt}=& \rho_{M} + \sum_{I}  \rho_{I}  + O(\epsilon^4) \label{emtt} \ ,  \\[5pt]
T_{t\mu} =& - \rho_{M} v_{M\mu} - \sum_{I} ( \rho_{I} + p_{I}) v_{I\mu} + O(\epsilon^5) \label{emtx} \ , \\[5pt]
T_{\mu \nu} =& \sum_{I} p_{I} \delta _{\mu\nu} + O(\epsilon^4) \ , \label{em}
\end{eqnarray}
where $\rho_I \sim p_I \sim \epsilon^2$ and $v_I \sim \epsilon$. In Ref. \cite{vaas2} we applied the post-Newtonian expansion to fluids of this type and found that energy-momentum conservation implies
\begin{equation} \label{emcon1}
\nabla_{\mu} \, p_I = 0 +O(\epsilon^4) \, .
\end{equation}
We therefore have that $p_I=p_I (t)\sim \epsilon^2$ is a function of time only, and not a function of space. For a barotropic fluid with equation of state $\rho_I=\rho_I (p_I)$ this means that we also have $\rho_I=\rho_I (t)$, at $O(\epsilon^2)$. This further restricts the form of $v_I$ to correspond to the velocity field of a uniformly expanding fluid, as we will explain further in Section \ref{cosmo}. The reader may note that nothing in this description relies on any specific theory of gravity - only on the conservation of energy-momentum. For further details on barotropic fluids with $p \sim \rho$ in post-Newtonian expansions, the reader is referred to Ref. \cite{vaas2}.

\subsection{Additional potentials}
\label{apot}

The extra fluids described above, and the extra degrees of freedom that generically exist in modified theories of gravity, require additional gravitational potentials to be included in equations (\ref{alpha}) and (\ref{gamma}). We define these potentials implicitly through the Poisson equations
\begin{eqnarray}
\nabla^2 \Phi \equiv  -4\pi G \alpha \rho_{M} + \alpha_{c} +O(\epsilon^4)\ ,  \label{Phigen} \\[5pt] 
\nabla^2 \Psi \equiv -4\pi G \gamma\rho_{M} + \gamma_{c} +O(\epsilon^4) \ , \label{Psigen}
\end{eqnarray}
where $\Phi$ and $\Psi$ are the metric perturbations from equation \eref{weakfield}, and where $\{\alpha, \gamma, \alpha_{c},\gamma_{c} \}$ are a set of parameters (to be constrained by observation and experiment). The first two of these are $O(\epsilon^0)$, as before. The last two are of $O(\epsilon^2)$, and are constants in space. We intend these extra two parameters to include all sources for gravitational fields that are independent of position, including the barotropic fluids discussed above and the additional degrees of freedom that occur in modified theories of gravity\footnote{The reason why extra potentials are required for the extra gravitational degrees of freedom in cosmology will become clear when we consider examples, in Section \ref{examples}.}.

This choice of parameterization for the gravitational potentials $\Phi$ and $\Psi$ is motivated by (i) the fact that the potentials that appear in the PPN framework can be expressed as a hierarchy of Poisson equations, and (ii) the fact that Poisson equations and Gauss' divergence theorem guarantee that cosmological back-reaction will be small. The first of these points means that our extended framework will be able to encompass all theories that fit naturally into the PPN formalism. This includes an array of simple scalar-tensor, vector-tensor, and bi-metric theories of gravity \cite{Will}. The second point comes from the fact that the large-scale cosmological behaviour can be obtained by integrating the weak-field gravitational equations over small regions of space \cite{Tim1,vaas1,vaas2}. It is only if these equations are of the form given in (\ref{Phigen}) and (\ref{Psigen}) that Gauss' theorem can be used to link the rate of cosmic expansion to energy density in a straightforward way \cite{pierre1}. This will become clearer when we derive the effective Friedmann equations in Section \ref{cosmo}.

Let us now turn to considering the solutions to equations (\ref{Phigen}) and (\ref{Psigen}). In the standard approach to the PPN formalism one would use the Green's function for an asymptotically flat space, to write the solution to
\begin{eqnarray} \nonumber
\nabla^2 U \equiv  -4\pi G \rho_{M}  \qquad {\rm as} \qquad U= \int \frac{\rho(x^{\prime})}{\vert {\mathbf x} - {\mathbf x^{\prime} \vert}} d^3 x^{\prime} \, .
\end{eqnarray}
However, in the present case, where we wish to consider cosmology, this is not the appropriate solution. There are no asymptotically flat regions in cosmology, and so one must use a different Green's function. For the case of a large number of polyhedral regions of space, with the vanishing extrinsic curvature condition used on the boundary, we find that the relevant solution instead takes the more complicated form \cite{vaas1} 
\begin{eqnarray} \label{109}
U &= \bar{U}+ 4\pi G \int_{\Omega} \mathcal{G} \rho_{M} \, dV - 4\pi G M \int_{\partial \Omega} \frac{\mathcal{G}}{A} \, dA \, , 
\end{eqnarray} 
where $\bar{U}$ is the average value of the potential, $\mathcal{G}$ is a Green's function explained in Ref. \cite{vaas1}, $dA$ is a surface area element of the polyhedral region of space, and $A$ is the total surface area of the polyhedron. The derivation of this result required use of both Gauss' theorem and Neumann boundary conditions. We will not go into any further details of these solutions here. For more information the reader is referred to Ref. \cite{vaas1}, where explicit expressions for $\mathcal{G}$ are found for cubic lattice cells.

\subsection{Additional time dependence}

Finally, we must consider how the additional degrees of freedom from modified theories of gravity should be expected to behave in our new formalism, and what this means for the PPN parameters. For a theory with a scalar field, for example, the expansion is given in (\ref{s1}). In the standard approach to the PPN formalism one would assume $\bar{\phi}$ to be effectively constant, and only varying over cosmological time-scales (if at all). When considering gravity in the solar system these variations are entirely negligible. When considering modified gravity in cosmology, however, they are not. We therefore cannot neglect the time dependence of $\bar{\phi}$ in scalar-tensor theories. Similarly, we cannot neglect the time-dependence of $\bar{A}_t$ in vector-tensor theories, when we expand the extra vector field as in equations (\ref{v1}) and (\ref{v2}). As the values of the PPN parameters depend on these quantities, this means we also have to allow the PPN parameters to be functions of time, so that we have 
$$\alpha=\alpha(t) \, , \quad \gamma=\gamma(t) \, , \quad \alpha_c=\alpha_c(t) \,  \quad {\rm and} \quad \gamma_c=\gamma_c(t) \, .$$ 
This does not alter the functional form of the solutions to equations (\ref{Phigen}) and (\ref{Psigen}) in space, as they are still Poisson equations, but it does add an extra degree of time dependence to the source functions. This means that Gauss' theorem can still be used to derive the sources for the Friedmann equations, and that back-reaction can be expected to be small. Spatial dependence of the parameters above would ruin this result, and would not produce a Newtonian gravitational field on small scales. This will be explained further in Section \ref{cosmo}, and explicit example theories will be used to illustrate these points in Section \ref{examples}.

\section{A parameterized approach to cosmology} \label{cosmo}

Let us now put together the emergent expansion considered in Section \ref{sec2} and the effective field equations considered in Section \ref{sec3}. This will allow us to obtain a set of effective Friedmann equations without using any particular set of field equations. It will also allow us to present parameterized, consistent expressions for both the large-scale expansion, and the quasi-static limit of first-order cosmological perturbations, in terms of our extended set of PPN parameters.

\subsection{Conservation equations}
 
We will first derive conservation equations for each of the matter fluids using the energy-momentum conservation equation, $ T^{ab}_{\ \ ; a} =0$. Assuming that to leading order each fluid is non-interacting, we obtain the result in (\ref{emcon1}) from the $O(\epsilon^2)$ part of the Euler equation. At next-to-leading order we find \cite{vaas2}
\begin{eqnarray}
\rho_{M,t} + \nabla \cdot ({\rho_{M} \bm v}_{M} )= 0 + O(\epsilon^5) \, ,  \label{emcon3}
\\[5pt] 
\rho_{I,t} + (\rho_{I} + p_{I}) \nabla \cdot {\bm v}_{I} = 0 + O(\epsilon^5) \, ,  \label{emcon2}
\end{eqnarray}
where subscript $M$ again refers to non-relativistic pressureless matter, and subscript $I$ corresponds to the barotropic fluids with pressure at $O(\epsilon^2)$. {The assumption that fluids are not interacting at leading order gives standard dark energy models, with interactions expected to occur at higher orders. One could potentially also consider more exotic interacting dark energy models with interactions at leading order, but have chosen to neglect this possibility here.}

To integrate these equations we make use of Reynold's transport theorem, which for any space-time function $f$ gives
\begin{eqnarray}
\frac{d}{dt} \int_{\Omega} f \ dV = \int_{\Omega} f_{,t} \ dV + \int_{\partial \Omega} f \bm{v} \cdot d\bm{A} \, .
\end{eqnarray}
Integrating equation (\ref{emcon3}) over our small region of space, and then using Gauss' theorem and Reynold's theorem, therefore gives
\begin{equation} \label{dM}
\frac{d}{dt} \int_{\Omega} \rho_M dV \equiv \frac{dM}{dt} = 0 \, ,
\end{equation}
where the first equality defines $M$. This means that $\langle \rho_M \rangle = M/V$, where the angle brackets denote the average value of $\rho_M$ in the spatial domain $\Omega$, and $V$ is the spatial volume of $\Omega$. In terms of the expanding coordinate system, equation (\ref{dM}) can be written as
\begin{eqnarray} \label{emcon4}
\boxed{\langle \rho_M \rangle_{,t}  + 3 \frac{\dot{a}}{a} \langle \rho_M \rangle=0} \, ,
\end{eqnarray}
which is, of course, just the usual conservation equation for dust in an FLRW space-time.

To derive a conservation equation for the barotropic fluid in (\ref{emcon2}) we do not need to integrate it over space, as we have already found it to be homogeneously distributed (to leading order). If instead we simply note that a homogeneous fluid comoving with the boundaries of our region of space must have $v_I^a=u^a$, where $u^a$ is the time-like 4-vector field from equation (\ref{ua}), then this gives $\nabla \cdot \bm{v_{I}} = 3 \dot{a}/a$. Substituting into equation (\ref{emcon2}) then gives
\begin{eqnarray}
\boxed{\rho_{I,t}  + 3 \frac{\dot{a}}{a} ({\rho_{I}} + p_{I} )=0} \, ,  \label{continuity1}
\end{eqnarray}
which is, of course, identical to the FLRW continuity equation for such a fluid. The conservation laws for the leading-order parts of both the non-relativstic and the barotropic fluid are therefore unaltered from the homogeneous and isotropic case, even though we have allowed for extremely large density constrasts. These results depend on energy-momentum conservation, but are otherwise independent of the theory of gravity under consideration.
 
\subsection{Background expansion}

Our next task it to write the emergent expansion, discussed in Section \ref{emergent}, in terms of the parameters and quantites from Section \ref{apot}. Let us start by integrating the constraint equation (\ref{con1}) over the spatial domain, $\Omega$. The spatial curvature term in this equation can be written
\begin{equation}
R^{(3)}= \frac{6 k}{a^2} - \frac{4}{a^2} \hat{\nabla}^2 \hat{\Psi} +O( \epsilon^4) \, ,
\end{equation}
where we have chosen to use the expanding coordinates from equation (\ref{FLRW}). Integrating this quantity over $\Omega$, and using Gauss' theorem, then gives
\begin{equation}
\label{con1a}
\int_{\Omega}  R^{(3)} dV = \frac{6 k}{a^2} V - \frac{4}{a^2} \int_{\partial \Omega} \hat{\nabla} \hat{\Psi} \cdot d \bm{A} = \frac{6 k}{a^2} V \, ,
\end{equation}
where in the last equality we have used the result that extrinsically flat boundaries are totally geodesic, implying $\bm{n} \cdot \hat{\nabla} \hat{\Psi}\vert_{\partial \Omega} =0$ \cite{Tim1, eisen}. If we now consider the other term on the right-hand side of equation (\ref{con1}), and similarly integrate this over $\Omega$ then we get
\begin{equation}
\label{con1b}
\int_{\Omega} \nabla^2 \Psi \ dV = -4 \pi G \gamma \langle \rho_M \rangle V + \gamma_c V \, ,
\end{equation}
where we have used equation (\ref{Psigen}). Note that if either $\gamma$ or $\gamma_c$ had been functions of space, then the right-hand side of this equation would have been considerably more complicated. Putting equations (\ref{con1a}) and (\ref{con1b}) together with equation (\ref{con1}) then gives
\begin{eqnarray}
\boxed{\frac{\dot{a}^2}{a^2} =  \frac{8\pi G \gamma}{3}\avg{\rho_{M}} - \frac{2\gamma_{c} }{3}- \frac{k}{a^2}} \ , \label{fincon}
\end{eqnarray}
where we have written the left-hand side in terms of the quantities in the expanding coordinates, and divided through by $V$. This equation has exactly the same form as the first Friedmann equation of FLRW cosmology. It has, however, been derived without reference to the field equations, using only (an extended version of) the PPN metric.

Let us now derive an evolution equation. If we integrate equation (\ref{X1}) over $\partial \Omega$, and use Gauss' theorem, then we get
\begin{equation}
\int_{\partial \Omega} X_{,tt} dA = - 4 \pi G  \alpha \langle \rho_M \rangle V + \alpha_c V \, .
\end{equation}
This equation can be simplified further by noting that $X_{,tt}$ must be constant over $\partial \Omega$, in order for equation (\ref{X2}) to remain valid. We therefore have
\begin{eqnarray} 
\boxed{\frac{\ddot{a}}{a} = -\frac{4\pi G \alpha}{3}\avg{\rho_{M}} + \frac{\alpha_{c} }{3}} \ , \label{accgen2}
\end{eqnarray}
where we have divided through by $V$, written the left-hand side in terms of the quantities used in the expanding coordinate system, and used the fact that $A/V=3/X$ for regular convex polyhedra. This equation is identical to the second Friedmann equation, but has again been derived without recourse to the field equations. The reader may again note that the right-hand side of this equation would have been considerably more complicated if either $\alpha$ or $\alpha_c$ had been functions of space.

By using the conservation equation \eref{emcon4}, the constraint equation \eref{fincon}, and the acceleration equation \eref{accgen2}, we can derive one further constraint for this system. This can be found by differentiating equation \eref{fincon}, and is given by
\begin{equation}
{4\pi G\avg{\rho_{M}}  = \bigg(\alpha_{c} + 2\gamma_{c} + \frac{d \gamma_{c}}{d \ln a} \bigg)\bigg/ \bigg(\alpha - \gamma + \frac{d\gamma}{d \ln a} \bigg)}\, .\label{addcon1}
\end{equation}
The existence of this constraint means that the first and second Friedmann equations, \eref{fincon} and \eref{accgen2}, can be written entirely in term of the set of parameters $\{ \alpha, \gamma, \alpha_c, \gamma_c \}$.
 
\subsection{First-order perturbations}

Finally, let us consider the small-scale, first-order cosmological perturbations that arise within this framework. Using the transformations from equations \eref{phitran} and \eref{psitran}, the Poisson equations \eref{Phigen} and \eref{Psigen} transform to give
\begin{eqnarray}
\label{np1}
\boxed{\hat{\nabla}^2 \hat{\Phi} =  -4\pi G a^2 \alpha \delta\rho}  \ ,  \\[5pt]
\boxed{\hat{\nabla}^2 \hat{\Psi} = -4\pi G a^2\gamma \delta\rho} \ , \label{np2}
\end{eqnarray}
where $\hat{\nabla}^2 =\hat{\partial}_{\mu} \hat{\partial}_{\mu}$, and where $\delta \rho = \hat{\rho} - \avg{\rho_{M}}$. These are exactly the type of equations that one would expect to describe cosmological perturbations on small scales, in the quasi-static limit. The often considered gravitational constant parameter, $\mu$, and gravitational slip parameter, $\zeta$, can then be written in terms $\alpha$ and $\gamma$ as
\begin{eqnarray}
\mu \equiv -\frac{{\nabla}^2 \hat{\Psi}}{4\pi G a^2 \delta\rho}  =\gamma  \qquad {\rm and} \qquad  
\zeta \equiv \frac{\hat{\Psi} - \hat{\Phi} }{\hat{\Psi}} = 1 - \frac{\alpha}{\gamma} \, .
\end{eqnarray}
These expressions provide a direct link between the parameters used to test gravity in cosmology ($\mu$ and $\zeta$), and those used in weak-field slow-motion world of post-Newtonian gravity ($\alpha$ and $\gamma$).

We can now see that equations (\ref{emcon4}), (\ref{continuity1}), (\ref{fincon}), (\ref{accgen2}), (\ref{np1}) and (\ref{np2}) provide a consistent set of equations to evolve both the cosmological background, and first-order cosmological perturbations in the quasi-static limit. This is all given in terms of a set of four parameters $\{ \alpha, \gamma , \alpha_c ,\gamma_c\}$ that are functions of time only, and that can be directly related to the PPN parameters. We refer to this framework as ``parameterized post-Newtonian cosmology'' (PPNC). In the next section we will illustrate how our four parameters can be determined in some simple classes of dark energy models, and modified theories of gravity. Such relations will allow observational constraints on $\{ \alpha, \gamma , \alpha_c ,\gamma_c\}$ to be imposed on the parameters that appear in each of these theories.

Before moving on, let us now provide some {\it a posteriori} justification for why $\{ \alpha, \gamma , \alpha_c ,\gamma_c\}$ should be functions of time only. From the derivation of equations (\ref{fincon}) and (\ref{accgen2}) one can immediately see that any spatial dependence in either $\alpha$ or $\gamma$ would have resulted in sources proportional to $\langle \gamma \rho_M \rangle$ and $\langle \alpha \rho_M \rangle$ in the emergent Friedmann equations. This would mean that any situation where $\alpha$ or $\gamma$ have spatial dependence should be expected to result in strong cosmological back-reaction, so that the formation of structure would have a large effect on the background expansion. {This is because $\alpha$ and $\gamma$ are expected to be related to the local distribution of mass. The integrated quantities $\langle \gamma \rho_M \rangle$ and $\langle \alpha \rho_M \rangle$ would therefore be non-linear functions, and their precise value would depend on how matter is clustered. Spatial dependence of this type would modify the standard dust-like terms in the Friedmann equations.}  {So, while one would still have a consistent cosmology, the precise rate of expansion would no longer be insensitive to the distribution of the mass of objects.} This would make the use of FLRW solutions, as a model to interpret observations, questionable, at best.

Furthermore, if $\alpha_c$ or $\gamma_c$ had spatial dependence, then equations (\ref{np1}) and (\ref{np2}) would have had an additional source on their right-hand sides. This would mean that observations used to interpret $\hat{\Phi}$ and $\hat{\Psi}$ may not be directly linked to the mass density, and that one could (for example) have lensing of light in a situation where the matter is perfectly homogeneous. None of these outcomes are desirable, and it seems to us that they can only be avoided if $\{ \alpha, \gamma , \alpha_c ,\gamma_c\}$ do not vary in space. We will see in the following section that simple dark energy models and conservative theories of modified gravity do, in fact, obey these expectations.

\section{Worked examples} \label{examples}

In this section will investigate how specific example theories of gravity can be incorporated into the formalism described above. For each theory we will calculate the value of the set of parameters $\{ \alpha, \gamma, \alpha_c, \gamma_c\}$, using the weak-field and slow-motion limit of the theory. We will then use the method outlined in Section \ref{sec2} to determine the emergent cosmological expansion for each theory, by using the appropriate set of junction conditions. This will give a set of Friedmann-like equations that govern the emergent cosmological expansion, and which can be compared to the analogous equations that one finds when considering the actual FLRW solutions for each of the theories under consideration. The purpose of this is two-fold. Firstly, it shows that the method used in Section \ref{cosmo} does faithfully represent the perturbed Friedmann solutions of a wide class of modified theories of gravity. Secondly, it confirms that the effect of non-linear structure on the large-scale properties of the cosmology can be neglected at leading order in perturbation theory. This latter property is required if we are to make any sensible link between weak-field gravity and FLRW cosmology.

Our first worked example will be general dark energy models in Einstein's theory. As sub-cases of this we look at simple quintessence dark energy models with a minimally coupled scalar field, as well as the standard $\Lambda$CDM model. We then consider scalar-tensor and vector-tensor theories of gravity as further worked examples. These two classes of theories require additional junction conditions for the additional degrees of freedom that they contain. This is the case because theories in which the field equations contain at most second-order derivatives of the fundamental fields should generically be expected to obey junction conditions that imply the smoothness and continuity of each of these fields. For Einstein's theory, this just corresponds to equations (\ref{metjunc1}) and (\ref{metjunc2}), as the metric is the only dynamical degree of freedom in the theory. For modified theories of gravity, the extra degrees of freedom must satisfy a similar set of conditions.

\subsection{Dark energy models}

Let us first consider a general dark energy model where a dark fluid is minimally coupled to the metric. The gravitational theory in this case is still given by  Einstein's field equations,
 \begin{eqnarray}
R_{ab} &= 8\pi G \left( T_{ab} - \frac{1}{2} T g_{ab} \right) \, , 
\end{eqnarray}
where $T_{ab}= T_{M ab} +T_{I ab}$, and $T_{M ab}$ and $T_{I ab}$ are the energy-momentum tensors of non-relativistic matter and the dark fluid, respectively.
Using the metric from equation \eref{weakfield}, the Poisson equations we obtain for the gravitational potentials in the weak-field limit are then given by
\begin{eqnarray}
\nabla^2 \Phi  =  - 4\pi G\rho_{M}  - 4\pi G(\rho_{I} + 3p_{I}) \, , \label{de1} \\
\nabla^2 \Psi = - 4\pi G \rho_{M} - 4\pi G \rho_{I} \, .  \label{de2}
\end{eqnarray}
This immediately gives the PPN parameters as
\begin{eqnarray}
\alpha = \gamma =1 \, ,
\end{eqnarray}
which are, of course, the usual values of these parameters in Einstein's theory. Whenever $\alpha = \gamma =1$ we can use equations \eref{fincon}, \eref{accgen2} and \eref{addcon1} to find the consistency relations
\begin{eqnarray}
\alpha_{c} + 2\gamma_{c} + \frac{d \gamma_{c}}{d \ln a} = 0 \label{grcon} \ , \\ 
2\alpha_{c} - 2\gamma_{c}  = 6 \dot{H} + 9H^2 +  \frac{3k}{a^2} \ ,
\end{eqnarray}
where $H = \dot{a} /{a}$ is the Hubble rate. These equations must be obeyed by both $\alpha_c$ and $\gamma_c$. For the field equations given in (\ref{de1}) and (\ref{de2}) we find
\begin{eqnarray}
\alpha_{c} = - 4 \pi G (\rho_{I} + 3 p_{I}) \, , \ \\
\gamma_{c} = -4\pi G \rho_{I} \ .
\end{eqnarray}
equations \eref{fincon} and \eref{accgen2} can then be used to write
\begin{eqnarray} 
\frac{\dot{a}^2}{a^2} + \frac{k}{a^2} =  \frac{8\pi G \gamma}{3}\avg{\rho_{M}} + \frac{8\pi G  }{3}\rho_I \ , \\[5pt]
\frac{\ddot{a}}{a} = -\frac{4\pi G \alpha}{3}\avg{\rho_{M}} - \frac{4 \pi G}{3}  \left( \rho_I + 3 p_I \right) \ .
\end{eqnarray}
These are identical to the equations for an FLRW solution to Einstein's equations with a barotropic fluid. The consistency between these equations and the FLRW equations of the same theory shows that our PPNC construction works for general relativity with general barotropic fluids. 

If we specialize further, to the case of a quintessence field \cite{quintessence}, the we have that the energy density and pressure are given by $\rho_{I} = \frac{1}{2} \dot{\phi}^2 + V(\phi)$ and $p_{I} = \frac{1}{2} \dot{\phi}^2 - V(\phi)$, {where $\dot{\phi} = d \phi / d \hat{t} \sim O(\epsilon)$ and $V(\phi) \sim O(\epsilon^2)$}. This gives
\begin{eqnarray}
\alpha_{c} = - 8 \pi G \left( \dot{\phi}^2 -  V(\phi) \right) \, , \ \\
\gamma_{c} = -4\pi G \bigg(\frac{1}{2} \dot{\phi}^2 + V(\phi)\bigg) \ ,
\end{eqnarray}
where $\phi$ is the minimally-coupled scalar field and $V(\phi)$ is the potential of that field. We can now use equations \eref{fincon} and \eref{accgen2} to write the emergent cosmological expansion as
\begin{eqnarray} 
\frac{\dot{a}^2}{a^2} + \frac{k}{a^2} =  \frac{8\pi G \gamma}{3}\avg{\rho_{M}} + \frac{8\pi G  }{3}\bigg(\frac{1}{2} \dot{\phi}^2 + V(\phi) \bigg)\ , \\[5pt]
\frac{\ddot{a}}{a} = -\frac{4\pi G \alpha}{3}\avg{\rho_{M}} - \frac{8 \pi G}{3}  \left( \dot{\phi}^2 -  V(\phi) \right) \ .
\end{eqnarray}
These are again identical to the equations for an FLRW solution to Einstein's equations with a minimally coupled quintessence field. The only extra equation we get in this case is the propagation equation for the scalar field:
\begin{eqnarray} 
\ddot{\phi} = - 3 \frac{\dot{a}}{a}\dot{\phi} -\frac{dV(\phi)}{d\phi} \ ,
\end{eqnarray}
which can be derived from the continuity equation \eref{continuity1}. This shows our parameterization is consistent with quintessence models of dark energy. It must therefore also be consistent with the $\Lambda$CDM model, as this just correponds to the case where both $\phi$ and $V(\phi)$ are constant. In this case we can set $\Lambda = 8\pi G V(\phi)$, and our parameters reduce to $\alpha_{c} = \Lambda$ and $\gamma_{c} =-\frac{\Lambda}{2}$. The acceleration and constraint equations then reduce to the Friedmann equations of $\Lambda$CDM universe. Our parameterization therefore also works for the standard $\Lambda$CDM model.

\subsection{Scalar-tensor theories of gravity}

Let us now turn our attention to a general class of scalar-tensor theories of gravity. These theories are some of the simplest generic modifications that one can make to general relativity, and involve the addition of only one non-minimally coupled scalar field, $\phi$. In order to fit into the formalism above, we choose to work in the Jordan frame where energy-momentum is covariantly conserved. It then immediately follows that the worldlines of test particles are geodesic \cite{Will, Tim3}. The Lagrangian for the class of theories we wish to consider is given by
\begin{equation} \label{Lst}
L =\frac{1}{16\pi G}\bigg[\phi R - \frac{\omega(\phi)}{\phi} g^{ab}\phi_{; a} \phi_{; b} - 2\phi\Lambda(\phi)\bigg] + L_{m}(\psi, g _{ab}) \ ,
\end{equation}
so that the effective gravitational constant $G_{\rm eff}$, as determined by local weak-field experiments, is modified by the space-time varying scalar field $\phi(t, x^{\mu})$. The semicolons denote covariant derivative with respect to the metric $g_{ab}$, and $\omega(\phi)$ and $\Lambda(\phi)$ are general functions of $\phi$. Finally, $\psi$ denotes matter fields. This class of theories reduces to Brans-Dicke theory when $\Lambda=0$ and $\omega$ is a constant \cite{Brans}. We recover a $\Lambda$CDM model when $\omega \to \infty$, $\omega' /\omega^2 \to 0$ and $\Lambda$ is a constant.

The field equations can be determined from the Lagrangian in (\ref{Lst}) using variational principles, and can be manipulated into the form
\begin{eqnarray}
\hspace{-20pt} \phi R_{ab}  =  8\pi G \left( T_{ab} - \frac{1}{2} g_{a b} T \right) +  g_{a b} \bigg( \frac{1}{2} \square \phi + \phi\Lambda(\phi) \bigg) + \frac{\omega(\phi)}{\phi} \phi_{; a} \phi_{; b} + \phi_{; a b} \, ,\label{Riccifieldscalar}
\end{eqnarray}
with a propagation equation for the scalar field given by
\begin{equation}
\hspace{-20pt} (2\omega(\phi) +3 )  \square \phi = 8\pi G T - \omega'(\phi) g^{cd}\phi_{; c} \phi_{; d}  - 2\phi\Lambda(\phi) + 2\phi^2\Lambda'(\phi) \, . \label{phimatter}
\end{equation}
In these equations $T_{ab} = T_{M ab} + \sum_{I} T_{I ab}$ is the sum of the energy-momentum tensors of the non-relativistic matter and any non-interacting barotropic fluids that may be present. We have also written $\omega'(\phi) = d\omega(\phi)/d\phi$ and $\Lambda'(\phi) = d\Lambda(\phi)/d\phi$, and used $\square$ to denote the covariant d'Alembertian operator.

The first thing to do, when considering the post-Newtonian limit of these theories, is to expand the scalar field $\phi$. We do this in the following way
\begin{equation}
\phi = \bar{\phi} + \delta \phi + O(\epsilon^4) \, ,
\end{equation}
where $\bar{\phi} \sim \epsilon^0$ and $\delta \phi \sim \epsilon^2$. This is so far the same as the treatment of this field in the PPN formalism. However, we now note that the lowest-order field equations give
\begin{equation}
\bar{\phi}_{,\alpha} = 0 \qquad {\rm or, equivalently,} \qquad \bar{\phi}=\bar{\phi}(t) \, .
\end{equation}
This means that the lowest-order part of $\phi$ can be dependent on time, but not on spatial position. At this point in the standard PPN formalism one assumes that $\bar{\phi}$ is effectively constant (i.e. not varying in space or time). While this is likely to be a very good approximation in the Solar System, it is unlikely to be valid on the scales we wish to consider in cosmology. Indeed, we will find that we must allow $\bar{\phi}$ to be a function of time in order for the emergent cosmological expansion to match the behaviour predicted by the Friedmann equations. From now on we will refer to $\bar{\phi}(t)$ as the ``background'' value of the scalar field, and we will suppress its argument. The perturbation $\delta \phi=\delta \phi (x^\alpha,t)$ is dependent on both position in space and time, as usual.

Using the weak-field metric from equation \eref{weakfield}, and the field equations \eref{Riccifieldscalar}-\eref{phimatter}, we can now write a set of Poisson equations for the gravitational potentials. They are given by equations of the form given in (\ref{Phigen}) and (\ref{Psigen}), with the parameter values
\begin{eqnarray}
\alpha(t) =\bigg(\frac{2\omega + 4}{2 \omega + 3}\bigg)  \frac{1}{ \bar{\phi}}\, , \label{alphascalartensor}\\[5pt]
\gamma(t) = \bigg(\frac{2\omega + 2}{2 \omega + 3}\bigg) \frac{1}{ \bar{\phi}}\, . \label{gammascalartensor}
\end{eqnarray}
These are exactly the same expression that one derives in the standard PPN formalism \cite{Will}, except that they are now functions of time. The fact that local gravitational experiments determine the present day value of Newton's constant to be given by $G$ then requires
\begin{equation}
\alpha (t_0) = 1 \qquad {\rm or, equivalently,} \qquad \bar{\phi}(t_0) = \bigg(\frac{2\omega + 4}{2 \omega + 3}\bigg) \, ,
\end{equation}
where $t_0$ denotes the present time. This provides a boundary condition on the function $\alpha(t)$, which is now generically expected to be non-constant in time. It also allows us to write the present day value of $\gamma$ as
\begin{equation}
\gamma(t_0) =  \bigg(\frac{\omega + 1}{\omega + 2}\bigg) \, ,
\end{equation}
which is the usual value used in post-Newtonian gravitational experiments. One may note that in our case this is only a boundary condition on $\gamma(t)$, which is also generically expected to be a non-constant function of time.

From equations (\ref{Phigen}) and (\ref{Psigen}) we can also read off the values of the cosmological parameters $\alpha_c$ and $\gamma_c$. These are given by
\begin{eqnarray}
\hspace{-20pt}\alpha_{c}(t) &=& - \bigg(\frac{2\omega + 4}{2 \omega + 3}\bigg) \sum_{I} \frac{4\pi G\rho_{I}}{\bar{\phi}} + \bigg(\frac{2\omega + 2}{2 \omega + 3}\bigg)\bigg(-\sum_{I}\frac{12\pi Gp_{I}}{\bar{\phi}} + \Lambda(\bar{\phi}) \bigg)  \nonumber \\ &&\qquad - \frac{\omega(\bar{\phi})}{\bar{\phi}^2}\dot{\bar{\phi}}^2 - \frac{\ddot{\bar{\phi}}}{\bar{\phi}} 
+ \bigg( \frac{1}{2\omega + 3} \bigg) \left(\frac{\omega' \dot{\bar{\phi}}^2}{2\bar{\phi}} + \bar{\phi}\Lambda' (\bar{\phi})\right) \, , \\[5pt]
\hspace{-20pt} \gamma_{c}(t) &=& -\bigg(\frac{2\omega + 2}{2 \omega + 3}\bigg) \sum_{I} \frac{4\pi G\rho_{I}}{\bar{\phi}} -\bigg(\frac{1}{4 \omega + 6}\bigg)\bigg(\sum_{I}\frac{24\pi G p_{I}}{\bar{\phi}}  + (2 \omega+1) \Lambda(\bar{\phi}) \bigg) \nonumber \\
 &&\qquad - \frac{\omega(\bar{\phi})}{4\bar{\phi}^2}\dot{\bar{\phi}}^2 - \frac{\ddot{\bar{\phi}}}{2\bar{\phi}}  -\bigg(\frac{1}{2 \omega + 3}\bigg)\bigg(\frac{\omega'}{2\bar{\phi}}\dot{\bar{\phi}}^2+ \bar{\phi} \Lambda'(\bar{\phi}) \bigg) \, . \label{gcST}
\end{eqnarray}
These equations have no counterparts in the standard PPN formalism, as they are neglected in that case. However, it can be seen that if $\bar{\phi}$ is a function of $t$, or if barotropic fluids of a scalar field potential are present, then they are not equal to zero. They can also not be neglected on cosmological scales, as we will see below. Finally, one may note that in this case the potential $\Lambda(\bar{\phi}) \sim O(\epsilon^2)$ is not the same as a non-interacting fluid with $p_{I} = -\rho_{I}$.

The only other weak-field equation in this theory, other than equations (\ref{Phigen}) and (\ref{Psigen}), is the propagation equation for the scalar field. This is given by
\begin{eqnarray}
\hspace{-50pt} \nabla^2 \delta \phi = \frac{1}{2\omega + 3}\bigg(\omega' \dot{\bar{\phi}}^2 -8\pi \rho_{M}  -8\pi \sum_{I} (\rho_{I} - 3p_{I})  - 2\bar{\phi}\Lambda(\bar{\phi}) + 2\bar{\phi}^2\Lambda'(\bar{\phi})\bigg) + \ddot{\bar{\phi}} \, . \label{nabladeltaphi}
\end{eqnarray}
{One may note that the terms responsible for screening mechanisms are absent at this order, due to the post-Newtonian expansion we have deployed. They should, however, be expected to appear at higher orders.} In order to determine the cosmological equations, we now need to know the appropriate junction conditions for $\phi$. These are given by
\begin{eqnarray}
\bigg[\phi \bigg]^{(+)}_{(-)} =&0 \,  \label{scalarjunc} \qquad {\rm and} \qquad
\bigg[\mathcal{L}_{n} \phi \bigg]^{(+)}_{(-)} = 0 \, ,
\end{eqnarray}
which ensure the smoothness and continuity of the scalar field $\phi$ at the boundary of the region of space we are considering. For the extrinsically flat boundaries we consider here, these equations give $\mathcal{L}_{n} \phi = 0$, which can be expanded to obtain
\begin{eqnarray}
\mathbf{n} \cdot \nabla \delta \phi |_{x=X} = -  \dot{a}  \dot{\bar{\phi}} \hat{X}_{0} +O(\epsilon^4) \, , \label{scalarjuncfin}
\end{eqnarray}
where $\hat{X}_{0}$ is the constant position of the boundary in the expanding coordinate system. 

Integrating equations (\ref{Phigen}), (\ref{Psigen}) and (\ref{nabladeltaphi}) over our region of space, using Gauss' theorem and equation (\ref{scalarjuncfin}), then gives the cosmological expansion equations for a general scalar-tensor theory of gravity. These are given by
\begin{eqnarray}
\hspace{-20pt}\frac{\dot{a}^2}{a^2} +\frac{k}{a^2}  =  \frac{8\pi G}{3\bar{\phi}}\avg{\rho_{M}} +\frac{8\pi G }{3\bar{\phi}}\sum_{I}\rho_{I}  + \frac{\omega(\bar{\phi})}{6\bar{\phi}^2}\dot{\bar{\phi}}^2 -  \frac{\dot{\bar{\phi}} \dot{a}}{\bar{\phi} a} + \frac{\Lambda(\bar{\phi})}{3} \, ,  \label{confinscalar}
 \end{eqnarray}
and
\begin{eqnarray}
\hspace{-20pt}\frac{\ddot{a}}{a}  =  -\bigg(\frac{\omega+3}{6 \omega + 9}\bigg)\frac{8\pi G }{\bar{\phi}}\avg{\rho_{M}}  -\bigg(\frac{\omega+3}{6 \omega + 9}\bigg)\frac{8\pi G }{\bar{\phi}}\sum_{I}\rho_{I}  -\frac{8\pi G}{\bar{\phi}}\sum_{I}p_{I} \bigg(\frac{\omega}{2\omega + 3}\bigg)  \nonumber \\
\qquad - \frac{\omega(\bar{\phi})}{3\bar{\phi}^2}\dot{\bar{\phi}}^2 +  \frac{\dot{\bar{\phi}} \dot{a}}{\bar{\phi} a}  + \Lambda(\bar{\phi}) \bigg(\frac{2\omega}{6 \omega + 9}\bigg) + \frac{1}{2\omega + 3}\bigg( \frac{\omega'}{2\bar{\phi}}\dot{\bar{\phi}}^2 +\Lambda'(\bar{\phi}) \bigg) \, , \label{accfinscalar}
\end{eqnarray}
and
\begin{eqnarray}
\hspace{-50pt} \frac{\ddot{\bar{\phi}}}{\bar{\phi}} = \frac{1}{2\omega + 3}\bigg(\frac{8\pi G}{\bar{\phi}}\bigg(\avg{\rho_{M}} + \sum_{I} (\rho_{I} - 3p_{I})\bigg) -\frac{\omega' \dot{\bar{\phi}}^2}{\bar{\phi}} + 2\Lambda(\bar{\phi}) - 2\bar{\phi} \Lambda'(\bar{\phi})\bigg) - 3 \frac{\dot{a} \dot{\bar{\phi}}}{a \bar{\phi}}  \, . \label{STacc}
\end{eqnarray}
Equations \eref{confinscalar}-\eref{STacc} are identical to the standard FLRW equations we expect to obtain for scalar-tensor theories of gravity \cite{Tim3, stcheck1}, as well as corresponding precisely to the parameterized equations \eref{fincon} and \eref{accgen2}. The corresponding first-order quasi-static cosmological perturbations are also given precisely by equations (\ref{np1}) and (\ref{np2}), with $\alpha$ and $\gamma$ given by equations (\ref{alphascalartensor}) and (\ref{gammascalartensor}). {One may note that at this order of approximation, and with the assumptions we have made, we find no Yukawa potentials. Again, the terms responsible for these in massive scalar-tensor theories should be expected to appear at higher orders.}

This shows our parameterization produces both the correct cosmological expansion, and the correct first-order perturbations, for this class of scalar-tensor theories of gravity. It also shows that the parameterized framework presented in Section \ref{cosmo} is a very compact way of presenting the cosmological dynamics.

\subsection{Vector-tensor theories of gravity}

In this subsection we will consider a general class of vector-tensor theories of gravity. These theories have a  time-like vector field, $A^{a}$, that is non-minimally coupled to gravity, and whose evolution equations are linear and at most second order in derivatives \cite{Will}. Their Lagrangian is given by \cite{Nord2,Nord3,Nord1}
\begin{equation}
\hspace{-50pt} L =\frac{1}{16\pi G}\bigg[ R + \omega A_{a}A^{a} R + \eta A^{a} A^{b} R_{ab} - \epsilon F^{ab} F_{ab} + \tau A_{a;b} A^{a;b} \bigg] + L_{m}(\psi, g _{ab}) \, , \label{SVL}
\end{equation}
where $A^{a}$ is a dynamical time-like vector field, and the 2-form $F_{ab}$ is defined by $F_{ab} \equiv A_{b;a} - A_{a;b}$. The parameters $\omega, \eta, \epsilon$ and $\tau$ in this Lagrangian are all constants, and $\psi$ denotes the matter fields present in the theory. We could also have included a term dependent on $A_{\mu} A^{\mu}$ in (\ref{SVL}), but this would behave in the same way as the $\Lambda (\phi)$ term in scalar-tensor theories of gravity and would needlessly complicate the situation.

When the action obtained from equation (\ref{SVL}) is varied with respect to the metric, the field equations we obtain are given by
\begin{equation}
G_{ab} + \omega \Theta_{ab}^{(\omega)} + \eta\Theta_{ab}^{(\eta)} + \epsilon \Theta_{ab}^{(\epsilon)} + \tau\Theta_{ab}^{(\tau)}  = 8\pi GT_{ab} \ , \label{Gfieldvector}
\end{equation}
where $G_{ab} = R_{ab} - \frac{1}{2}g_{ab}R$ is the Einstein tensor, $T_{ab} = T_{M ab} + \sum_{I} T_{I ab}$ is the total energy-momentum tensor (including both matter and non-interacting fluids), and the $\Theta$'s are given by
\begin{eqnarray}
\hspace{-50pt} \Theta_{ab}^{(\omega)} &= A_{a} A_{b} R + A^2 R_{a b} - \frac{1}{2} g_{ab} A^2 R - (A^2)_{; ab} + g_{ab} (A^2)_{;c}^{\ ; c} \, , \label{SV1} \\
\hspace{-50pt} \Theta_{ab}^{(\eta)} &= 2 A^{c} A_{(a} R_{b) c} - \frac{1}{2} g_{ab} A^{c} A^{d} R_{c d} - (A^{c}A_{(a})_{;b)c} + \frac{1}{2} (A_{a} A_{b})_{;c}^{\ ; c} + \frac{1}{2} g_{ab} (A^{c} A^{d})_{; c d} \, ,   \\
\hspace{-50pt} \Theta_{ab}^{(\epsilon)} &= - 2(F^{c}_{\ a} F_{b c} - \frac{1}{4} g_{ab} F^{c d} F_{cd}) \, ,  \\
\hspace{-50pt} \Theta_{ab}^{(\tau)} &= A_{a;c} A_{b}^{\ ; c} + A_{c ; a} A^{c}_{\ ; b} - \frac{1}{2} g_{ab} A_{c;d} A^{c;d} + (A^{c} A_{(a;b)} - A^{c}_{; (a} A^{}_{b)} - A^{}_{(a} A_{b)}^{\ ; c} )_{; c} \, ,
\end{eqnarray}
where $A^2 = A^{a} A_{a}$. 
The field equation obtained by varying the action from equation (\ref{SVL}) with respect to the vector field $A_{a}$ is given by
\begin{equation}
\epsilon F^{ab}_{ \quad ; b} + \frac{1}{2} \tau A^{a ; b}_{ \quad ; b} - \frac{1}{2} \omega A^{a} R - \frac{1}{2} \eta A^{b} R^{a}_{\ b} = 0 \ . \label{Afield}
\end{equation}
The field equations (\ref{Gfieldvector}) - (\ref{Afield}) give the full set of field equations for the theories we wish to consider in this subsection.

Let us now expand the components of the vector field $A_{a}$, in the post-Newtonian limit. For this we write
\begin{eqnarray}
A_t = \bar{A}_{t} + \delta A_{t} + O(\epsilon^4) \, , \\
A_{\mu} = \delta A_{\mu} + O(\epsilon^3) \, ,
\end{eqnarray}
where $\bar{A}_{t} \sim \epsilon^0$, and $\delta A_{\mu} \sim \epsilon^1$, and $\delta A_{t} \sim \epsilon^2$. The reader may note that we have taken the leading-order perturbation to the spatial component of the vector field to contribute at $O(\epsilon)$, which differs from the standard treatment in the PPN formalism, where the lowest-order part of this component is usually taken to be $O(\epsilon^3)$. We find that this is necessary in order to reproduce the correct large-scale expansion. 

Using the field equations \eref{Gfieldvector} - \eref{Afield} we find that the leading-order part of the time component of the vector field must obey
\begin{equation}
A_{t,\alpha} =0 \qquad {\rm or, equivalently,} \qquad \bar{A}_{t} = \bar{A}_{t}(t) \, . 
\end{equation}
This also differs from the standard PPN formalism, which assumes that any time dependence in $A_{t}$ can be neglected at this order. Again, such an assumption is likely to be valid on small scales (such as in the Solar System), but will not generically be valid on cosmological scales. In fact, just as with the scalar field in the previous section, we find that we require $\bar{A}_t$ to be time dependent in order to reproduce the expected large-scale expansion. We will refer to $\bar{A}_t$ as the ``background'' value of $A_t$, and note that $\delta A_{t}$ is expected to be a function of both space and time.


Let us now consider the lowest-order field equations that feature $\delta A_{\mu}$. Using the $t\mu$-component of equation \eref{Gfieldvector} 
and the spatial component of equation \eref{Afield} we find
\begin{eqnarray}
\tau (\eta + \tau - 4\epsilon)\bar{A}_{t} \delta A_{\mu, \nu \nu} = 0 \, .
\end{eqnarray}
This means that if $\tau (\eta + \tau - 4\epsilon) \bar{A}_{t} \neq 0$ (as one should expect in general circumstances), then we must have $\delta A_{\mu, \nu \nu} =0$. We can then see that equation \eref{Afield} implies that $\delta A_{\mu, \mu \nu} =0$, which implies $\delta A_{\mu, \mu} = f(t)$ for some function $f(t)$. In general, the solution for $\delta A_{\mu}$ can therefore be written as
\begin{eqnarray}
\delta A_{x} = \frac{1}{3} f(t) x + C_{1}(t,y,z) \, , \label{vecsoln1} \\
\delta A_{y} = \frac{1}{3} f(t) y + C_{2}(t,x,z) \, , \\
\delta A_{z} = \frac{1}{3} f(t) z + C_{3}(t,x,y) \, ,
\end{eqnarray}
where $C_{1}$, $C_{2}$ and $C_{3}$ are unknown functions to be determined.

At this point it is useful to consider the junction conditions on the vector field $A_a$. For theories with at most two derivatives in the field equations we expect smoothness and continuity to imply the following:
\begin{eqnarray}
\bigg[A^{\parallel}_{i}\bigg]^{(+)}_{(-)} = 0 \label{vecjun1} \, , \qquad
\bigg[A^{\perp} \bigg]^{(+)}_{(-)} = 0 \, , 
\qquad {\rm and} \qquad
\bigg[ (\mathcal{L}_{n}A)_{i} \bigg]^{(+)}_{(-)}  = 0 \, , \label{vecjun3}
\end{eqnarray}
where $A^{\parallel}_{i} \equiv ({\partial x^{a}}/{\partial \xi^{i}}) A_a$ is the component of the vector field that is parallel to the boundary, where $A^{\perp} \equiv n^a A_a$ is the component of the vector field that is perpendicular to the boundary, and where $(\mathcal{L}_{n}A)_{i} \equiv ({\partial x^{a}}/{\partial \xi^{i}}) \mathcal{L}_{n}A_{a}$ is the Lie normal derivative of the vector field projected on the boundary. The $\xi^i$ here refer to a set of coordinates on the boundary of the region of space being considered.

Under reflection symmetric boundary conditions, the last two equations in \eref{vecjun3} simplify to $A^{\perp}=0$ and $(\mathcal{L}_{n}A)_{i}=0$. Then, using equations \eref{metjunc1}, \eref{metjunc2}, and \eref{vecjun3}, we find that the value of the $x$-component of the vector field on the boundary should be given by $\delta A_{x}|_{x=X} = - \dot{a} \bar{A}_{t} \hat{X}_{0}$, where $\hat{X}_{0}$ is a constant. From this and equation \eref{vecsoln1} we can infer that $f(t) = -3({\dot{a}}/{a}) \bar{A}_{t}$ and $C_{1}(t,y,z) = 0$. Similar considerations lead to the results $C_{2}(t,x,z) =C_{3}(t,x,y)= 0$, so that we end up with
\begin{eqnarray}
\delta A_{x} =& - \frac{\dot{a}}{a} \bar{A}_{t}  x \, , \\
\delta A_{y} =& - \frac{\dot{a}}{a} \bar{A}_{t}  y \, , \\ 
\delta A_{z} =& - \frac{\dot{a}}{a} \bar{A}_{t}  z \, . \label{zvector}
\end{eqnarray}
These results will be very useful for simplifying a lot of the terms that will occur in the equations below.

Using the weak-field metric from equation \eref{weakfield}, and the field equations \eref{Gfieldvector} - \eref{zvector}, we can now write another set of Poisson equations for the gravitational potentials in these theories. They are again given by equations of the form given in (\ref{Phigen}) and (\ref{Psigen}), with the parameter values
\begin{eqnarray}
\alpha =- \frac{1}{\mathcal{D}} \bigg[ 2 \omega  \bar{A}_{t}^2 ( \tau -8 \omega -2 \epsilon )+2   (2 \epsilon -\tau ) \bigg] \ , \label{alphavectortensor}\\[5pt]
\gamma = -\frac{1}{\mathcal{D}}\bigg[ 2  \omega \bar{A}_{t}^2   (-2 \eta +\tau -4 \omega +2 \epsilon )  +2  (2 \epsilon -\tau ) \bigg]\ ,  \label{gammavectortensor}
\end{eqnarray}
where $\mathcal{D}$ is a function of time, and is given by
\begin{eqnarray}
\hspace{-20pt} \mathcal {D} =& -\omega  \bar{A}_{t}^4 \left(-\eta ^2+4 \eta  \omega +\tau ^2-10 \tau  \omega +12 \omega ^2+4 \epsilon  (\eta -\tau +3 \omega )\right) \nonumber \\
&\qquad +\bar{A}_{t}^2 \left(-\eta ^2+4 \eta  \omega +\tau ^2-4 \tau  \omega +12 \omega ^2+4 \epsilon  (\eta -\tau )\right)+2 \tau -4 \epsilon \, .
\end{eqnarray}
These expressions for $\alpha$ and $\gamma$ are generally functions of time, but reduce to the usual expression in PPN gravity when the time dependence of $\bar{A}_t$ is neglected. As before, the fact that local gravity experiments measure the value of Newton's constant to be $G$ means that we have the boundary condition $\alpha(t_0)=1$, which gives the present day value of $\bar{A}_t=\bar{A}_t(t_0)$.

We can again read off the value of the cosmological parameters $\alpha_c$ and $\gamma_c$ from equations (\ref{Phigen}) and (\ref{Psigen}). These are still only functions of time, and are given by
\begin{eqnarray}
\fl \alpha_{c} = 
 \frac{1}{\mathcal{D}}\bigg[  8 \pi  G \sum_{I} \bigg(\omega \bar{A}_{t}^2 (3 p_{I} (-2 \eta+\tau -4 \omega +2 \epsilon )+\rho_{I}  (\tau -8 \omega -2 \epsilon ))
 +(3 p_{I}+\rho_{I} ) (2 \epsilon -\tau ) \bigg)\nonumber \\
     \fl \qquad \qquad -6  \bar{A}_{t}^2 \frac{\ddot{a}}{a} \bigg(\omega  \bar{A}_{t}^2 \bigg(-2 \eta ^2-4 \eta  \omega +\tau ^2-6 \tau  \omega +\epsilon  (3 \eta -\tau +6 \omega )\bigg) \nonumber \\ 
 \qquad \qquad \qquad -\tau (\eta +\tau )+\epsilon  (\eta +3 \tau +2 \omega )\bigg) \nonumber \\
   \fl \qquad \qquad -6 \bar{A}_{t} \dot{\bar{A}}_{t} \frac{ \dot{a}}{a}  \bigg(\omega  \bar{A}_{t}^2 (-(2 \eta +\tau ) (2\eta - \tau + 4\omega)+\epsilon  (5\eta +\tau + 6\omega ))\nonumber \\
 \qquad \qquad \qquad   -\tau  (2 \eta +\tau )+\epsilon  (3 \eta +3 \tau +2 \omega )\bigg) \nonumber \\
  \fl \qquad \qquad   -3 \bar{A}_{t}^2 \frac{\dot{a}^2}{a^2} (-\eta +2 \omega +2\epsilon ) \left(2 \omega  \bar{A}_{t}^2 (\eta +\tau )-\tau \right) \nonumber \\
   \fl \qquad \qquad +2  \bar{A}_{t}\ddot{\bar{A}}_{t} \bigg(\omega  \bar{A}_{t}^2 \left(3 \eta ^2-2 \eta  (\tau -6 \omega )+2\omega  (6 \omega -\tau )+\epsilon  (-3 \eta +\tau -6 \omega )\right) \nonumber \\
  \qquad \qquad \qquad - (\epsilon  (3 \eta +\tau +6 \omega )-2 \tau  (\eta +\omega))\bigg)  \nonumber \\ 
   \fl \qquad \qquad +\dot{\bar{A}}_{t}^2 \bigg(2 \omega  \bar{A}_{t}^2 \left(3 \eta ^2-3 \eta  \tau +12 \eta  \omega +\tau ^2-8 \tau  \omega +12 \omega ^2+\epsilon (-3 \eta +\tau -6 \omega )\right) \nonumber \\
  \qquad \qquad \qquad  + (2 \epsilon -\tau ) (-3 \eta +2 \tau -6 \omega )\bigg) \bigg] \, ,  
\end{eqnarray}
and
\begin{eqnarray}
\fl \gamma_{c} =\frac{1}{4\mathcal{D}}\bigg[  16 \pi  G  \sum_{I}\bigg(3 p_{I} \bar{A}_{t}^2 (\eta -\tau +2 \omega ) (-\eta -\tau -2 \omega +4 \epsilon )\nonumber \\
 \qquad \qquad \qquad  +2 \rho_{I} \bar{A}_{t}^2 \omega  (-2 \eta +\tau -4 \omega +2 \epsilon ) +2 \rho_{I}  (2 \epsilon -\tau ) \bigg)\nonumber \\
 \fl \qquad \qquad -6  \bar{A}_{t}^2 \frac{ \ddot{a}}{a} \bigg(\bar{A}_{t}^2 \bigg(-2 \eta ^3-\eta ^2 (\tau +8 \omega )+2 \eta  \left(\tau ^2-4 \tau  \omega -4 \omega ^2\right)\nonumber \\ 
 \qquad \qquad \qquad +\tau 
   \left(\tau ^2+2 \tau  \omega -12 \omega ^2\right)+4 \epsilon  \left(2 \eta ^2-\eta  \tau +7 \eta  \omega -\tau ^2+6 \omega ^2\right)\bigg) \nonumber \\
  \qquad \qquad \qquad -2\left(\tau ^2+2 \epsilon  (\eta -2 \tau +2 \omega )\right)\bigg) \nonumber \\
   \fl \qquad \qquad +12 \bar{A}_{t} \dot{\bar{A}}_{t} \frac{ \dot{a}}{a}  \bigg((\eta - \tau + 2\omega) (2\epsilon +\bar{A}_{t}^2((2\eta + \tau) (\eta + \tau + 2\omega) - 2\epsilon (4\eta + 2\tau + 3\omega) )\bigg)\nonumber \\
  \fl \qquad \qquad   -3 \bar{A}_{t}^2 \frac{\dot{a}^2}{a^2} \bigg(\bar{A}_{t}^2 \bigg(-2 \eta ^3-\eta ^2 \tau +2 \eta  (\tau -2 \omega )^2+\tau  \left(\tau ^2+6 \tau  \omega -4 \omega ^2\right) \nonumber \\
  \qquad \qquad \qquad +\epsilon  \left(8\eta ^2-4 \eta  (\tau -2 \omega )-4 \tau  (\tau +\omega )\right)\bigg) \nonumber\\
   \qquad \qquad \qquad  +2 (\tau  (4 \eta +\tau +4 \omega )-2 \epsilon  (2 \eta +3 \tau))\bigg) \nonumber \\
   \fl \qquad \qquad +2  \bar{A}_{t}\ddot{\bar{A}}_{t} \bigg(2\tau  (\eta +2 \omega -2 \epsilon )- \bar{A}_{t}^2 \bigg(-3 \eta ^3-18 \eta ^2 \omega  \nonumber \\
  \qquad \qquad \qquad +\eta \left(3 \tau^2+2 \tau  \omega -36 \omega ^2\right) +2 \omega  \left(\tau ^2+2 \tau  \omega -12 \omega ^2\right)\nonumber \\ 
   \qquad \qquad \qquad +4 \epsilon  \left(3 \eta ^2-3 \eta  (\tau -4\omega )+\omega  (12 \omega -5 \tau )\right)\bigg) \bigg) \nonumber \\
   \fl \qquad \qquad +\dot{\bar{A}}_{t}^2 \bigg( \bar{A}_{t}^2 \bigg(6 \eta ^3-3 \eta ^2 (\tau -12 \omega )-2 \eta \left(3 \tau ^2+8 \tau  \omega -36 \omega ^2\right) \nonumber \\
    \qquad \qquad \qquad   - 4 \epsilon  \left(6 \eta ^2-9\eta  \tau +24 \eta  \omega +3 \tau ^2-19 \tau  \omega +24 \omega ^2\right) \nonumber \\
    \qquad \qquad \qquad +3 \tau ^3-10 \tau ^2 \omega -20 \tau  \omega ^2+48 \omega ^3 \bigg)+2 \tau  (2 \epsilon -\tau )\bigg) \bigg] \ . \label{gammacvector}
\end{eqnarray}
Again, these equations do not exist in the standard PPN formalism, as time-dependence of the background fields is neglected in that case. However, it can be seen that if $\bar{A}_t$ is a function of $t$, or if barotropic fluids are present, then they are non-zero.

The final weak-field Poisson equation is the propagation equation for $\delta A_{t}$. This is given by
\begin{eqnarray}
\fl \nabla^2 \delta A_{t} =\frac{1}{\mathcal{D}} \bigg[ 8 \pi  G \rho_{M} \bigg(\omega  \bar{A}_{t}^3 (\eta -\tau +6 \omega )- \bar{A}_{t}  (\eta -\tau -2 \omega )\bigg)\nonumber \\
 +8 \pi  G  \sum_{I} \bigg( \omega  \bar{A}_{t}^3 (\rho_{I}  (\eta -\tau +6 \omega )+9 p_{I} (\eta -\tau +2 \omega )) \nonumber \\
\qquad \qquad \qquad - \bar{A}_{t} (\rho_{I}  (\eta -\tau -2 \omega )+3 p_{I} (\eta -\tau +2 \omega))\bigg) \nonumber \\
 +6 \bar{A}_{t}  \frac{\ddot{a}}{a} \bigg(\bar{A}_{t}^2 \left(\eta ^2+2 \eta  \omega -\tau ^2-2 \eta  \epsilon +2 \tau  \epsilon \right) + \omega\bar{A}_{t}^4 \left(-3 \eta^2+\eta  (\tau -6 \omega ) \right)\nonumber \\
 \qquad \qquad \qquad +\omega  \bar{A}_{t}^4 \left(2 \tau  (\tau -3 \omega )+2 \epsilon  (\eta -\tau +3 \omega )\right)+2 \epsilon \bigg) \nonumber \\
+6   \dot{\bar{A}}_{t}\frac{\dot{a}}{a}\bigg(\omega  \bar{A}_{t}^4 (2 \epsilon  (\eta -\tau +3 \omega )- 3(2 \eta +\tau ) (\eta -\tau +2 \omega )) \nonumber \\
 \qquad \qquad \qquad  -\bar{A}_{t}^2 (2 \epsilon  (\eta -\tau )-(2 \eta+\tau ) (\eta -\tau -2 \omega ))+2 \epsilon \bigg) \nonumber \\
 -3 \bar{A}_{t} \frac{\dot{a}^2}{a^2} \bigg(\omega  \bar{A}_{t}^4 \left(\eta ^2+4 \eta  \omega -\tau ^2-10 \tau  \omega+12 \omega ^2+4 \epsilon  (\eta -\tau +3 \omega )\right) \nonumber \\
  \qquad \qquad \qquad +\bar{A}_{t}^2 \left(\eta ^2-\eta  \tau -8 \eta  \omega -12 \omega ^2-4 \eta  \epsilon +4 \tau  \epsilon \right)+4 \epsilon \bigg) \nonumber \\
  +\ddot{\bar{A}}_{t} \bigg(-\omega  \bar{A}_{t}^4 \left(9 \eta ^2-10 \eta  \tau +36 \eta  \omega +\tau ^2-18 \tau  \omega+36 \omega ^2\right) \nonumber \\
   \qquad \qquad \qquad + \bar{A}_{t}^2 \left(3 \eta ^2-4 \eta  (\tau -3 \omega )+\tau ^2-8 \tau  \omega +12 \omega ^2\right)+2 \tau  \bigg) \nonumber \\
  +\dot{\bar{A}}_{t}^2 \bigg( \bar{A}_{t} \left(3 \eta ^2-5 \eta  \tau +12 \eta  \omega +2 \tau ^2-8 \tau  \omega +12 \omega ^2\right) \nonumber \\
   \qquad \qquad \qquad  -\omega  \bar{A}_{t}^3 \left(9 \eta ^2-14 \eta  \tau +36 \eta  \omega +5 \tau ^2-30 \tau  \omega +36 \omega ^2\right)\bigg) \bigg] \, . \label{pertfinal}
\end{eqnarray}
In this case, taking the time component of the last of the expressions in \eref{vecjun3} gives
\begin{eqnarray}
\mathbf{n} \cdot \nabla \delta A_{t} |_{x=X} =& \frac{\dot{a}^2}{a} \bar{A}_{t} \hat{X}_{0} - \dot{a} \dot{\bar{A}}_{t} \hat{X}_{0} - \ddot{a} \bar{A}_{t} \hat{X}_{0}\ . \label{vecjunfin}
\end{eqnarray}
Integrating equations (\ref{Phigen}), (\ref{Psigen}) and (\ref{pertfinal}) over our spatial domain, using Gauss' theorem and equation (\ref{vecjunfin}), then gives the equations for the cosmological evolution of the space-time. Firstly, the constraint equation in these theories is given by
\begin{eqnarray}
\hspace{-45pt} \frac{\dot{a}^2}{a^2} = -\frac{16 \pi G (\avg{\rho_{M}} + \sum_{I}\rho_{I}) a^2 + \tau a^2 \dot{\bar{A}}_{t}^2 + 6(\eta + 2\omega)\dot{a} a \bar{A}_{t}  \dot{\bar{A}}_{t} -6k(1-\omega \bar{A}_{t}^2)}{3 a^2 (-2 + (2\eta + \tau + 2\omega) \bar{A}_{t}^2)} \ . \label{confinvect}
\end{eqnarray}
Next, the acceleration equation is given by
\begin{eqnarray}
\fl \frac{\ddot{a}}{a} = \frac{8\pi G (\avg{\rho_{M}} + \sum_{I}\rho_{I}) (-2 \tau + (8\eta \tau + \tau ^2 - 12 \eta \omega + 14 \tau \omega - 24 \omega^2)\bar{A}_{t}^2)}{3 (-2 + (2\eta + \tau + 2\omega) \bar{A}_{t}^2) (-2\tau + (-3\eta^2  + \tau^2 + 2\eta(\tau - 6\omega) + 2 \tau \omega - 12 \omega^2)\bar{A}_{t}^2)} \nonumber \\[5pt]
\fl \qquad \ \ +  \frac{8 \pi G\tau \sum_{I}p_{I}}{  (-2\tau + (-3\eta^2 + \tau^2 + 2\eta(\tau - 6\omega) + 2 \tau \omega - 12 \omega^2)\bar{A}_{t}^2)} \nonumber \\[5pt]
\fl \qquad \ \ +    \frac{2\dot{\bar{A}}_{t}^2\tau(3\eta-2\tau + 6\omega + (-3\eta^2 + \tau^2 + 2\eta(\tau - 6\omega) + 2 \tau \omega - 12 \omega^2)\bar{A}_{t}^2)}{3 (-2 + (2\eta + \tau + 2\omega) \bar{A}_{t}^2) (-2\tau + (-3\eta^2 + \tau^2 + 2\eta(\tau - 6\omega) + 2 \tau \omega - 12 \omega^2)\bar{A}_{t}^2)} \nonumber \\[5pt]  
\fl \qquad \ \ +\frac{ 6k (\eta + 2\omega )\bar{A}_{t}^2 (2(\eta+ \tau) \omega \bar{A}_{t}^2 - \tau)}{a^2 (-2 + (2\eta + \tau + 2\omega) \bar{A}_{t}^2) (-2\tau + (-3\eta^2 + \tau^2 + 2\eta(\tau - 6\omega) + 2 \tau \omega - 12 \omega^2)\bar{A}_{t}^2)} \nonumber \\[5pt]  
\fl \qquad \ \ + \frac{\bar{A}_{t} \dot{\bar{A}}_{t} \dot{a}}{a} \frac{(4\omega-2\tau)}{(-2 +  (2\eta + \tau + 2\omega) \bar{A}_{t}^2)} \, . \label{accfinvect}
 \label{flrwacc}
\end{eqnarray}
Finally, the evolution equation for the background value of the vector-field is given by
\begin{eqnarray}
\hspace{-40pt} \frac{\ddot{\bar{A}}_{t}}{\bar{A}_{t}}=- \frac{8\pi G (\avg{\rho_{M}} + \sum_{I}\rho_{I})  (\eta + 2\tau - 2\omega) + 24\pi G \sum_{I}p_{I}  (\eta + 2\omega) }{(-2\tau + (-3\eta^2 + \tau^2 + 2\eta(\tau - 6\omega) + 2 \tau \omega - 12 \omega^2)\bar{A}_{t}^2)} -\frac{3 \dot{\bar{A}}_{t} \dot{a}}{\bar{A}_{t} a} \nonumber \\[5pt]
-  \dot{\bar{A}}_{t}^2  \frac{ (-3\eta^2 + \tau^2 + 2\eta(\tau - 6\omega) + 2 \tau \omega - 12 \omega^2)}{(-2\tau + (-3\eta^2 + \tau^2 + 2\eta(\tau - 6\omega) + 2 \tau \omega - 12 \omega^2)\bar{A}_{t}^2)}  \nonumber \\[5pt]
-k \frac{12 \omega  \bar{A}_{t}^2 (\eta +\tau ) - 6 \tau}{a^2(-2\tau + (-3\eta^2 + \tau^2 + 2\eta(\tau - 6\omega) + 2 \tau \omega - 12 \omega^2)\bar{A}_{t}^2)} \ .
\end{eqnarray}
These three equations are again identical to the Friedmann equations of this class of theories, showing that the emergent expansion proceeds as expected. They are also identical to the parameterized expressions presented in equations (\ref{fincon}) and (\ref{accgen2}), with the appropriate values of $\{\alpha,\gamma,\alpha_c,\gamma_c\}$. Once more, the first-order quasi-static cosmological perturbations are given by equations (\ref{np1}) and (\ref{np2}), this time with $\alpha$ and $\gamma$ given by equations (\ref{alphavectortensor}) and (\ref{gammavectortensor}). 

This shows that the parameterization we presented in Section \ref{cosmo} is again applicable, even though the equations are much more complicated in this case. This again highlights the highly compact nature of the parameterized expressions presented in Section \ref{cosmo}, and its ability to incorporate theories that fit into the PPN formalism.

\section{Conclusions}

In this paper we have constructed a parameterization that extends and transforms the PPN formalism for use in cosmology. This framework is not simply built in analogy to the PPN formalism, but is actually isometric to it on suitably defined spatial domains (that is, the two systems are actually equivalent in a physically meaningful sense). The result is a set of parameterized cosmologies that are fully consistent with the standard framework that is used to constrain gravity in the weak-field slow-motion limit of gravity, and that can be used to test Einstein's gravity and its many alternatives on cosmological scales. The advantage of this approach is that the consistency requirement with PPN requires that the parameters involved must be functions of time only. It also gives constraints on the present day values of some of these parameters, if local experiments are to measure the correct value of Newton's constant, $G$, and an experimentally acceptable value of the spatial curvature caused by rest mass, $\gamma$. If one did allow for spatial dependence in our parameters then the result would not be compatible with PPN, and should generically be expected to lead to either strong back-reaction or gravity without the presence of rest mass (depending on the parameter in question).

Formally, we end up with a generic system of Friedmann equations, and linear-order scalar perturbations in the quasi-static limit, that are valid for any theories of gravity that fit into the PPN approach. Our full set of parameters is given by the functions $\{ \alpha (t), \gamma (t), \alpha_{c} (t),\gamma_{c} (t) \}$. The first two of these reduce to the corresponding PPN parameters when $t=t_0$, and the second two are new ``cosmological'' parameters that determine the rate of expansion and acceleration in the large-scale cosmology. The correspondence with PPN parameters means that cosmological observations can be used to either (i) impose constraints on $\alpha$ and $\gamma$ over cosmologically interesting scales that complement those obtained from isolated astrophysical systems, or (ii) impose the following boundary conditions on the initial values of $\alpha$ and $\gamma$:
\begin{equation}
\alpha(t_0) = 1 \qquad {\rm and} \qquad \gamma(t_0) =   1 + (2.1 \pm 2.3) \times 10^{-5} \, .
\end{equation}
The former of these ensures that local gravitational experiments measure the correct value of $G$, and the latter is the experimentally determined value of $\gamma$ from observations of the Shapiro time-delay effect of radio signals from the Cassini spacecraft as they pass by the sun \cite{cassini}. In case (ii), observations at high redshifts could be used to impose constraints on the variation of $G$ as the Universe evolves, by constraining $\alpha(t)$ at times $0<t<t_0$.

Observationally, one can constrain the parameters $\{ \alpha (t), \gamma (t), \alpha_{c} (t),\gamma_{c} (t) \}$ with the cosmological probes that are, by now, quite standard in constraining modified theories of gravity. Importantly, however, we allow for the background expansion to be a part of the parameterization. This is required for most minimal modifications to Einstein's theory, including the scalar-tensor and vector-tensor theories considered in this paper, and offers new ways to constrain the underlying theory. We also have equation \eref{addcon1}, which provides a consistency relation between our parameters, and may reduce the number of observables required to constrain our full set of parameters. In terms of specific observables, one could for example use supernova data to constrain the Hubble rate $H=\dot{a}/a$ and the acceleration $\ddot{a}/a$ \cite{nova1, nova2, nova3, nova4, nova5, nova6}. Independent information on the density of baryons and dark matter (e.g. from primordial nucleosynthesis) together with information on the spatial curvature of the Universe (e.g. from CMB \cite{CMB} and BAO observations \cite{baos}), should then provide constraints on $\alpha_c(t)$ and $\gamma_c(t)$. Cosmological perturbations, on the other hand, can be used together with observations of the growth rate of structure to determine $\alpha(t)$, and together with observations of weak-lensing to determine the combination $\alpha(t)+\gamma(t)$. This is just a schematic of what is possible of course, and a large number of other cosmological probes are also available to provide additional constraints. In general, we expect there to be more observational probes than parameters in this framework, meaning that the system should be able to be constrained effectively with existing and upcoming data.

Of course, there are also certain limitations to our formalism. It does not, for example, apply to many of the more complicated theories of gravity that are now frequently considered in cosmology, as such theories do not always fit into the PPN framework. These theories may include Yukawa potentials \cite{fR, bimetric} or involve non-perturbative gravity \cite{Chameleon, Vainshtein} in the weak-field regime, both of which we have neglected to consider here. {We have also only been concerned with small-scale perturbations, in what is often referred to as the quasi-static limit of cosmological perturbation theory. The inclusion of large-scale perturbations is required to complete the picture, and these may lead to the presence of Yukawa potentials. }These subjects will be addressed in future studies, although it has already recently been shown that one should generically expect Yukawa potentials to lead to strong back-reaction \cite{pierre1}, and we strongly suspect the same applies to theories that involve non-perturbative screening mechanisms. Including more complicated theories, and large-scale perturbations, should therefore be expected to lead to significant complication in the parameterized framework. In this sense, one can consider the PPNC framework we have outlined here as a minimal construction for testing minimal deviations from Einstein's theory. This is sufficient to use tests of gravity from cosmology to constrain conservative theories, as is usual in both Solar System and binary pulsar applications of the PPN formalism.

\section*{Acknowledgements}
We are grateful to P. Carrilho, J. A. V. Kroon, P. Fleury and S. Imrith for helpful discussions and comments. VAAS and TC both acknowledge support from the STFC.

\end{document}